\documentclass[aps,prl,amsmath,amssymbb,reprint,superscriptaddress]{revtex4-2}

\usepackage{graphicx}% Include figure files
\usepackage{dcolumn}% Align table columns on decimal point
\usepackage{bm}% bold math
\usepackage[utf8]{inputenc}
\usepackage[T1]{fontenc}
\usepackage{graphicx}
\usepackage{enumitem}
\usepackage{amssymb}
\usepackage{hyperref}
\usepackage{amsthm}
\usepackage{tensor}
\usepackage{subcaption}
\usepackage{physics}
\usepackage{amsmath,amsfonts,amssymb}

\hypersetup{
   colorlinks,
    citecolor=blue,
    filecolor=black,
    linkcolor=blue,
    urlcolor=black
}

\newcommand{\smeq}{ \! = \!}

\newcommand{\tria}{{\tilde r}_{I,\alpha}}
\newcommand{\ria}{r_{I,\alpha}}
\newcommand{\ngamin}{{n^\gamma_{\alpha ,\rm min}}}

\usepackage[normalem]{ulem}

\begin{document}

\title{A Social structure description of epidemics propagation with the mean field game paradigm}

\author{Louis Bremaud}
\author{Denis Ullmo}
\affiliation{ Universit\'e Paris-Saclay, CNRS, LPTMS, 91405, Orsay, France}

\date{\today}

\begin{abstract}

We consider the spread of infectious diseases through a Mean Field Game version of a SIR compartmental model with social structure, in which individuals are grouped by their age class and interact together in different settings. In our game theoretical approach, individuals can choose to limit their contacts if the epidemic is too virulent, but this effort comes with a social cost. We further compare the Nash equilibrium obtained in this way with the societal optimum that would be obtained if a benevolent central planner could decide of the strategy of each individual, as well as to the more realistic situation where an approximation of this  optimum is reached through social policies such as lockdown.

\end{abstract}
\maketitle

%\section{Introduction}

As the Covid-19 has made rather explicit in the last few years, possessing good prediction tools for the dynamics of virus infections is mandatory if one wishes to design public policies making it possible to mitigate effectively the negative impact of an epidemic. Since the early twentieth century, many models have been proposed to address this issue, one of the simplest being the SIR (Susceptible-Infected-Recovered) compartment model \cite{SIR_Mckendrick} and its variations \cite{mathematics_infectious_diseases}, which has been  recently refined to take into account the structure of  social contacts \cite{Inferring_social_structure}, \cite{mistry2020inferring} or spatial/geographic aspects of the dynamics \cite{spatiotemporal_dynamics_epidemics}, \cite{Nature_social_networks}.

For virus epidemics like the Covid-19, with a very fast dynamics, one important difficulty met by epidemiologists can be illustrated already on the SIR model.  Noting $S$, $I$ and $R$ the relative proportion of agents in the three possible states (respectively ``Susceptible'', ``Infected'' and ``Recovered''), the time dependence of these ``state variables'' follow the set of equations
\begin{equation}
\label{eq:SIR}
\begin{cases}
\displaystyle{\Dot{S} = -  \chi S(t)I(t)} \\
\displaystyle{\Dot{I} = ( \chi S(t) -  \xi)I(t)} \; , \\
\displaystyle{\Dot{R} =   \xi I(t)} \\
\end{cases}
\end{equation}
which are characterized by two {\em ``extrinsic''} parameters, the recovery rate $\xi$ and the contact rate $\chi$.

Given the height of the stakes posed by the control of the Covid-19 epidemics in the last couple of years, both from a public health and economic point of view, major efforts have been invested by the epidemiologist community to extract these parameters, or their counterpart in more complex models, from the actual data observed on the field.  However, if $\xi$ is mainly fixed by biological considerations, and thus can be considered as essentially constant in time, the contact rate  $\chi$ on the other hand depends a lot on the agent's behavior (ie whether they actually meet or not) which has a dynamics on its own.  This dynamics is furthermore coupled to the dynamics of the epidemic itself since people will limit or increase their contacts depending on whether or not they feel at risk from the epidemic.  This implies that it is essentially impossible to fit the time dependence of $\chi$ on past data. In models used to advise public policies, this time dependence is thus either simply ignored, or involves a lot of guesswork, leading to predictions that can be trusted only for a rather short amount of time (see nevertheless \cite{Mobility_data_covid,Impact_NPI_Imperial}).
To avoid such a situation, it is necessary to introduce models  whose {\em ``extrinsic''} parameters have no time dependence (on the time scale of the epidemic), and which can therefore be fitted in a reliable way on field data.  In other words, it is necessary to make {\em ``intrinsic''} the dynamics of parameters such as $\chi$. To achieve this, a game theoretical approach is required,  and the one  that we will follow here is provided by Mean Field Game (MFG) theory. 

Introduced by Lasry and Lions a decade ago \cite{Lions_MFG, Lions_optimal_control, Lions_stationnaire} and independently by M.Huang, R.P. Malhamé and P. E. Caines \cite{Huang_Malhame_MFG}, Mean Field Games (MFG) focus on the derivation of a Nash equilibrium within a population containing a larger number of individuals.  Reader can look at \cite{Dynamic_Game_theory_Caines,continous_time_2013,Probabilistic_theory_MFG_Carmona} for a complete mathematical description, and to \cite{Ullmo_Quadratic_MFG, MFG_Ullmo_Shrodinger} for an introduction designed for physicists.  Applications of MFG, include finance \cite{Gueant_MFG_applications}, economics \cite{MFG_Bertrand_Cournot} and opinion dynamics \cite{Opinion_dynamics_MFG} among others.  
The introduction of MFG models to describe epidemics dynamics has been pioneered by Turinici and al. to describe vaccination strategies \cite{Laguzet_vaccination_SIR} or the dynamics of the parameter $\chi(t)$ in the simple SIR model \cite{Turinici_contact_rate_SIR_simple}.

The simple toy models addressed in \cite{Turinici_contact_rate_SIR_simple} are however presumably still too schematic to be relevant from a practical, public policy point of view.  The goals of this letter are to show that a good degree of complexity can be included in these MFG models, and in particular that we can implement a description of the social structure of society in which the epidemics develop.  Furthermore we shall see that with our Mean Field  Game approach, question of direct practical importance, such as defining the best government strategy with respect to confinement and deconfinement policies, can be addressed.

%\section{A social structure based modeling of the epidemic dynamics}

%\paragraph{A MFG social structure based Model}

We therefore consider a  SIR model with a structure of social contacts proposed in \cite{Inferring_social_structure} and \cite{mistry2020inferring} to get a more detailed description of the society, at a mesoscopic scale.  
Following \cite{Inferring_social_structure}, we make a differentiation between individuals according to their age. Here we choose to introduce three age classes : ``young'', ``adult'' and ``retired'' people but a more refined description could easily be implemented.  Furthermore, we split the society in four main settings where individuals have contacts with other : the schools, the households, the community and the workplaces. Thus the dynamics of the epidemic may differ for different age class and the interactions between individuals (of the same class or not) may differ in different setting. 

To model the interactions, following \cite{Inferring_social_structure}, we introduce the parameters $M^\gamma_{\alpha \beta}$ which measure the average frequency of contacts with someone of age class $\beta$ for an individual of age class $\alpha$ in the setting $\gamma$. To enforce the sum rule imposed by the fact that a contact between two agents involves both of them in a symmetric way, we make a slight variation here with respect to \cite{Inferring_social_structure} and set $M^\gamma_{\alpha \beta} = W^\gamma_{\alpha \beta} \cdot K_{\beta}$ where $W^\gamma_{\alpha \beta}$ is a symmetric matrix and $K_{\beta}$ is the proportion of individuals of age class $\beta$ in the population. Physically, $W^\gamma_{\alpha \beta}$ can be seen as the ``willingness of contacts'' between an individual of age class $\alpha$ and another of age class $\beta$ in the setting $\gamma$. We assume here that this symmetric matrix is built as  $W^\gamma_{\alpha \beta} = w^{\gamma }_{\alpha \beta} \cdot w^{\gamma}_{\beta \alpha}$ where $w^{\gamma}_{\alpha \beta}$ is the ``willingness'' of an individual of age class $\alpha$ to have contact with someone of age class $\beta$ (in the setting $\gamma$). 

In our game theoretical approach, we assume that individuals of age class $\alpha$ control their ``willingness of contacts'' with other individuals in each  setting.  We therefore write $w^{\gamma}_{\alpha \beta} = w^{\gamma (0)}_{\alpha \beta} n^{\gamma}_{\alpha}(t) $, where $w^{\gamma (0)}_{\alpha \beta}$ denotes this ``willingness'' in the absence of epidemic (similarly for $W^{\gamma (0)}_{\alpha \beta}$ and $M^{\gamma (0)}_{\alpha \beta}$), and  $n^{\gamma}_{\alpha}(t) \in [\ngamin,1]$ is a time dependent coefficient measuring the effort made by the individual to limit contacts because of the epidemic situation, and which is assumed to vary between a value $\ngamin$ representing the maximum effort that can be expected from the agent and 1 corresponding to the base willingness in the absence of effort. Notice that for simplicity, we use $n^{\gamma}_{\alpha}$ instead of $n^{\gamma}_{\alpha \beta}$, that is individuals do not change their behavior according to the age class of the contact $\beta$ but only according to the setting $\gamma$ (a $\beta$ dependence of $n$ could easily be implemented to this model and change only slightly the equations).

Indexing by $\alpha$ the proportion of Susceptible/ Infected/ Recovered people of age class $\alpha$, and  denoting by $q$ the probability of transmission (of the virus) per effective contact (between a susceptible and an infected), the SIR equations  (with $n=3$ age classes) read \cite{Inferring_social_structure}
\begin{equation}
\label{eq:SIR-ss}
\begin{aligned}
\dot S_\alpha & = -   \bar{\lambda}_\alpha (t) S_\alpha(t) \; \\
\dot I_\alpha & = +   \bar{\lambda}_\alpha (t)   S_\alpha(t) -  \xi I_\alpha(t) \; ,\\
\dot R_\alpha & =   \xi I_\alpha(t)
\end{aligned}
\end{equation}
where the ``force of infection'' $\bar{\lambda}_\alpha (t)$  corresponds to $q$ time the average number of infected people met by a susceptible agent of age class $\alpha$ during $dt$, and is written as 

\begin{equation}
\label{eq:lambda}
   \bar{\lambda}_\alpha (t) \equiv   q \left [ \sum_{\beta=1}^n \sum_{\gamma} \ {\bar n}^\gamma_\alpha (t) \ {\bar n}^\gamma_\beta (t)  \ M^{\gamma(0)}_{\alpha \beta} \ I_\beta(t) \right ] \; , 
\end{equation}
with $\bar{n}^\gamma_\alpha$ the average value of ${n^\gamma_\alpha}$ over agents in the age class $\alpha$. In the following, we will denote $\lambda$ (without bar above) when we focus on the force of infection seen by a reference individual, $\lambda_\alpha (t) \equiv   q \left [ \sum_{\beta=1}^n \sum_{\gamma}  { n}^\gamma_\alpha (t)  {\bar n}^\gamma_\beta (t)  M^{\gamma(0)}_{\alpha \beta}  I_\beta(t) \right ] $. 

In our Mean Field Game version of this model, the state variable of an agent $k_\alpha$ of age class $\alpha$ is her status $x_{k_\alpha} \in \{ s_\alpha \smeq {\rm susceptible}, i_\alpha \smeq {\rm infected}, r_\alpha \smeq {\rm recovered} \}$. The control parameters of individuals of age class $\alpha$ are the contact willingness $n^\gamma_\alpha(t)$, and each  individual ${k_\alpha}$ which is susceptible at time $t$ (ie $x_{k_\alpha} = s_\alpha$) will adjust these contact willingness to minimize an inter-temporal cost that we take of the form
\begin{align} 
    &  C_\alpha \left(\{n^\gamma_\alpha (\cdot)\},t\right)  \equiv \label{eq:cost} \\
    & \int_t^{T}   \left[ \lambda_\alpha(s) 
      \tria(I(s))
     +  f_\alpha \left (\{n^\gamma_\alpha(s)\} \right ) \right ]   
    (1 - \phi_\alpha^I(s)) ds \; . \nonumber
\end{align}
In this equation
\begin{equation} \label{eq:total-cost}
    \tria(I(s)) = \ria + g_\alpha (I(s))
\end{equation}
is the total cost of infection, which includes a base cost $\ria$ (which we assume increasing with the age class $\alpha$, modeling that we suffer more from infection when we are older), and an additional cost $g_{\alpha}(I(s))$ which model the saturation of the sanitary system. $f_\alpha \left (\{n^\gamma_\alpha (s)\} \right )$ measure the cost (both psychological and financial) associated with the limitation of social contacts (we assume this cost to be decreasing, with a positive second derivative), and  $\phi_\alpha^I(t)$ is the probability for an individual of age class $\alpha$ to be infected before $t$, so that an infection for this individual happens between $t$ and $t+dt$ with a probability $(1 - \phi_\alpha^I(t)) \lambda_\alpha(t) dt $. Note that in principle one should also model specifically the behavior of infected people, as this could vary from a completely egoistic approach where they stop doing any effort  to a very altruistic one where infected people completely isolate from the rest of population.  In epidemics like covid-19, however, most of the transmission is due to a small part of the infected people not aware of their infectious status.  Our model correspond to the limit where this proportion is extremely small, and for which $q$, the probability of transmission of the virus, integrate this probability.

To solve this optimization problem, we follow a standard approach in this context \cite{continous_time_2013}, and introduce the {\em value function}
\begin{equation}
\label{eq:value function}
    U_\alpha(t) = \underset{\{n^\gamma_\alpha(\cdot) \}}{\min} C_\alpha\left(\{n^\gamma_\alpha(\cdot) \},t\right) 
\end{equation} 
which is thus the minimal price (in stochastic average) that a susceptible agent (at $t$) can pay between $t$ and the end of the game.  Using the Bellman equation, which states that, for any intermediate time $t_i$, the optimal path between $t$ and $T$ can be constructed as the concatenation of optimal paths between $t$ and $t_i$ and between $t_i$ and $T$ followed by an optimization of the control parameters {\em at} $t_i$  we get the  Hamilton-Jacobi-Bellman equation of our Mean Field Game 
\begin{equation} \label{eq:HJB}
- \frac{dU_\alpha(t)}{dt} = \\ \underset{\{n^\gamma_\alpha(t) \}}{\min} \left [ \lambda_\alpha(t) \left ( \tria (I(t)) - U_\alpha(t) \right ) + f_\alpha \left (\{n^\gamma_\alpha(t)\} \right ) \right ]   \; .
\end{equation}
Then, the optimal strategy $n^{\gamma*}_\alpha(t)$ is expressed as 
\begin{equation} \label{eq:nstar}
\{n^{\gamma*}_\alpha(t)\} = \\  \underset{\{n^\gamma_\alpha(t)\}}{\text{argmin}} \left [\lambda_\alpha(t) \left ( \tria (I(t)) - U_\alpha(t) \right ) + f_\alpha \left (\{n^\gamma_\alpha(t)\} \right ) \right ] \; .
\end{equation}

We stress, however, that in Eq.~\eqref{eq:cost}, the dynamic of the infection Eqs.~\eqref{eq:SIR-ss}-\eqref{eq:lambda} at time $t$, is fixed by the strategies $\bar n^\gamma_\alpha(s<t)$, followed (on average) by the total population of agents, which is a priory distinct from the one $n^{\gamma*}_\alpha$ followed by the  individual optimizing the cost Eq.~\eqref{eq:cost}.  In all rigor this cost should be written as $C_\alpha\left(\{\bar n^\gamma_\alpha \} , \{n^\gamma_\alpha \},t\right)$, and the situation for which for all setting $\gamma$ and all age class $\alpha$ one has 
\begin{equation} \label{eq:Nash}
    n^{\gamma *}_\alpha = \bar{n}^\gamma_\alpha
\end{equation} 
corresponds to a Nash equilibrium, in the sens that an individual agent has no interest in deviating to another strategy if this strategy is followed by the rest of the agents.  ``Solving'' our Mean Field Game therefore amounts to~: i) Solve the rate equations \eqref{eq:SIR-ss} assuming the general population strategy $\{\bar n^\gamma_\alpha \}$ given. This in particular will determine the function $\phi_\alpha^I(t)$ for $t \in [0,T]$;  ii) Solve the optimization problem defined by the cost Eq.~\eqref{eq:cost} and deduce from it $\{ n^{\gamma*}_\alpha \}$, the optimal $\{ n^\gamma_\alpha \}$ for a given individual; and iii) Impose the self consistent equation \eqref{eq:Nash} that defines the Nash equilibrium of our MFG.
   
Since the time dependence of the $\{n^{\gamma}_\alpha \}$ is now an outcome of the description, our MFG model defined by the dynamics Eqs.~\eqref{eq:SIR-ss}-\eqref{eq:lambda} and the cost function Eq.~\eqref{eq:cost} clearly meet the criterion that all the extrinsic parameters characterizing it are time independent, and could in principle be fitted on field data. The actual extraction of these parameters is of course well beyond the scope of this work, and in the following, we illustrate the behavior of our MFG for a ``reasonable choice'' of this parametrization (these quantities are rather generic, and the observed behaviors are a priori typical, which was checked by running many simulations with different parameters).

As mentioned above, we consider four settings ($S$=schools, $W$=workplaces, $C$=community and $H$=households) and three age classes ($y$=youth, $a$=adults, $r$=retired). For the cost of infection Eq.~\eqref{eq:total-cost} we take
\begin{equation}
\label{eq:cost_infection}
    \tria \left( I(t) \right) = \kappa_\alpha \left[r_I 
    \left ( \exp\left[\alpha_{\rm sat} \frac{I(t)-I_{\rm sat}}{I_{\rm sat}}\right]  -1 \right) \right] \; ,
\end{equation} 
where the factors $\kappa_\alpha$ accounts for the fact that older agents are more impacted by the infection,  while $r_I$ and $\alpha_{\rm sat}$ are both constant modeling respectively the usual cost of infection and the impact of saturation on the cost. The additional cost is exponential with a threshold when we reach the saturation at $I = I_{\rm sat}$. Finally for the cost of the contact willingness reduction $f_\alpha(\{n^\gamma_\alpha\})$  we take a form inspired from \cite{Turinici_contact_rate_SIR_simple}
\begin{equation}
\label{eq:falpha}
    f_\alpha(\{n^\gamma_\alpha(t)\}) = \sum_{\gamma} \left [\left (\frac{1}{n^\gamma_\alpha(t)} \right )^{\mu_\gamma} - 1 \right ] \; ,
\end{equation} 
where $\mu_\gamma $ models= the variability of the ``attachment''  to the setting $\gamma$, as it is for example easier to reduce contacts at work rather than inside families.

\begin{table*}
    \centering
    \begin{tabular}{ c c c c c c c c c  }
     \hline 
     $M^S$ & $M^W$ & $M^C$ & $M^H$ & $\kappa_\alpha$ & $K_\alpha$ & $n^\gamma_{min}$ &  $\mu_{\gamma}$ \\ \hline
      $\begin{pmatrix}
    100 & 0 & 0 \\
    0 & 0 & 0 \\
    0 & 0 & 0
    \end{pmatrix}$  & 
    $\begin{pmatrix}
    0 & 0 & 0 \\
    0 & 75 & 0 \\
    0 & 0 & 0
    \end{pmatrix}$ & $\begin{pmatrix}
    12.5 & 25 & 12.5 \\
    12.5 & 25 & 12.5 \\
    12.5 & 25 & 12.5
    \end{pmatrix}$ & $\begin{pmatrix}
    15 & 25 & 10 \\
    12.5 & 32.5 & 5 \\
    10 & 10 & 30
    \end{pmatrix}$  & $(1,10,1000)$ & $(0.25, 0.5, 0.25)$ &  $(\frac{1}{3},\frac{1}{5},\frac{1}{5},\frac{1}{2})$ & $(2,2,1,3)$ \\ \hline 
    
    $S_\alpha(0)$ & $I_\alpha(0)$ & $(I_{sat},\alpha_{sat})$  & $\gamma$ & $q$ & $I_l$ & $I_d$ & $\sigma$ \\ \hline
    $(0.99,0.99,0.99) $ & $(0.01, 0.01, 0.01)$  & (0.01,0.01) & 1.2 & 0.02 & 0.1 & $4.10^{-4}$ & 0.39 \\ \hline 
    \end{tabular}
    \caption{Table of parameters used in our simulations (proportion in \% for epidemics quantities, the time unit is assumed to be a week, matrices and rates have the corresponding units)}
    \label{table:M}
\end{table*}

\begin{figure}[htb]
\begin{center}
\includegraphics[scale=0.42]{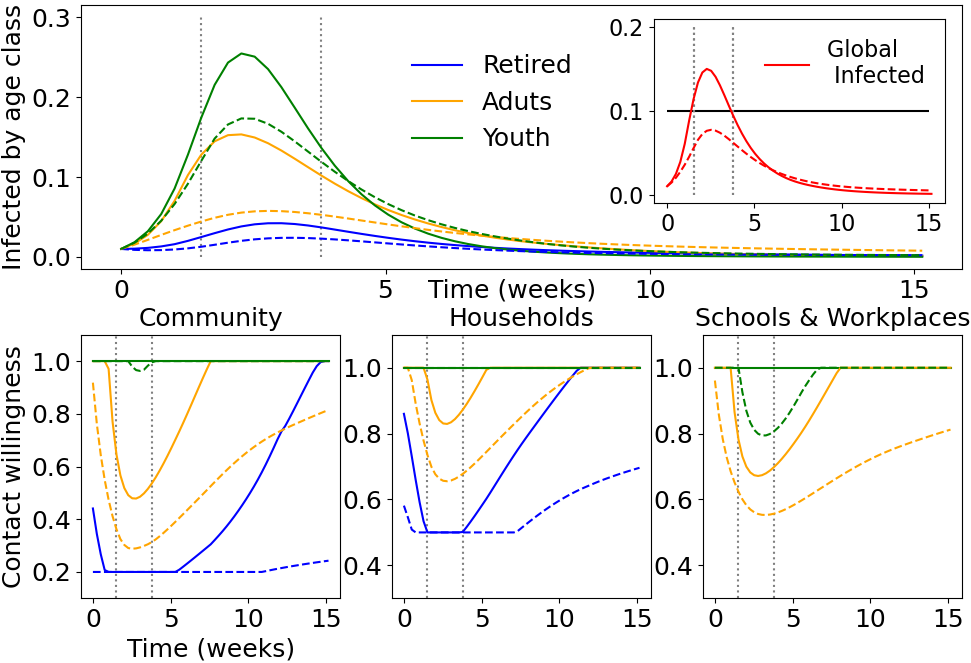}
\end{center}
\caption{Evolution of the epidemic quantities and contact willingness with $r_I = 1$ (solid line) and $r_I = 5$ (dashed line). Upper panel~: evolution of proportion of infected by age class. Lower panel (left to right)~: evolution of contact willingness of individuals according to their age class in community, households, schools (for the young) and workplaces (for the adults)}
\label{fig.MFG-free}
\end{figure}

Fig.~\ref{fig.MFG-free} shows the dynamics of the epidemic together with the choices made  by individuals for their contact willingness for both a relatively moderate cost for the infection ($r_I = 1$) and a much stronger one ($r_I=5$), with the choice of parameters given in table~\ref{table:M}.  The simulations have been obtained using a gradient descent on the variable $\{n_\alpha^\gamma \}$ of the cost $C$ to reach the Nash equilibrium. In the case $r_I = 1$, we see in this figure that there are significant efforts made by individuals when $I(t)$ exceed the threshold $I_{\rm sat}$. More precisely, retired people significantly reduce their contacts because the cost associated with  the infection is for them very high and this reduction is done in particular in the community setting because this is the easiest place to reduce once contacts. On the other hand, young people, who take no significant risk with the disease, barely modify their behavior, while the adults are in an intermediate situation. For $r_I = 5$, the cost of infection is sufficiently high so that one does not reach the saturation $I_{\rm sat}$, the epidemic is lower and slower.
%here

\begin{figure}[htb]
\begin{center}
\includegraphics[scale=0.41]{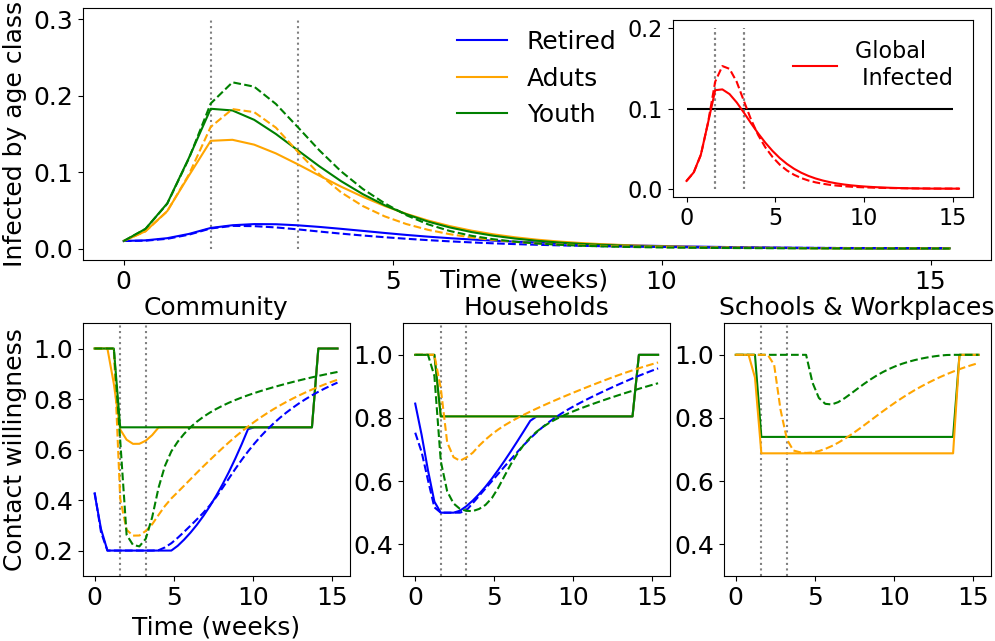}
\caption{Evolution of the epidemic quantities and contact willingness for the societal optimum (dashed line) and the optimal lockdown policy  (solid line). Upper panel : evolution of proportion of infected by age class (main panel) and on average (inset). Lower panel (left to right) : evolution of contacts willingness of individuals according to their age class in community, households, schools, and workplaces.}
\label{fig.SO_constraints}
\end{center}
\end{figure}

In the previous equilibrium analysis, each agent performs a personal, eventually egoistic, optimization. A ``benevolent global planner'', ie a well meaning government with full empowerment, would, on the other hand, attempt to reach a ``societal optimum'' \cite{optimal_isolation_policies,morton_wickwire_1974,Turinici_contact_rate_SIR_simple}, i.e.\  to optimize the global cost of the entire society, which would amount to solve : 
\begin{equation}
\label{eq:societal_optimum}
\underset{\{\bar{n}^\gamma \}}{\min} \; C_{\rm glob} \left(\{\bar{n}^\gamma_\alpha \} \right) \; \equiv \; \underset{\{\bar{n}^\gamma \}}{\min} \; \sum_{\alpha} \left [K_\alpha \times C_\alpha \left(\{\bar{n}^\gamma_\alpha \},\{\bar{n}^\gamma_\alpha \} \right) \right ]  \; \;  .
\end{equation}
The difference between this new minimization and the Nash equilibrium discussed above is referred to as  ``the cost of anarchy'', because there is no cooperation between individuals in the Nash equilibrium contrary to the societal optimum case. We compute it with a gradient descent on the cost $C_{glob}$, and we plot the dynamics on figure ~\ref{fig.SO_constraints}.

In practice however, it is largely impossible for a government to control the detailed behavior of each individual, especially in democratic countries, and even if this was technically feasible, it would involve an important coordination cost that would have to be included in the epidemic cost Eq.~\eqref{eq:cost}.  Government will therefore use median mode of actions, such as lockdown, to approach the societal optimum at a reasonable coordination (and democratic) cost.  We address now the question of how the lockdown policy can be used to approach as well as possible the societal optimum.

We therefore assume that above a certain threshold of infection,  $I_l$, a global planner impose a reduction of the maximum contact willingness $n^{\gamma}_{\alpha l}$, that we assume of the form $n^\gamma_{\alpha l} = \sigma n^{\gamma}_{\alpha ,{\rm min}} + (1-\sigma)$, ($\sigma \in [0,1]$) in each setting for each individual. Once the {\em ``lockdown''} is imposed, one returns to the original situation (without lockdown) at different threshold $I_d$ with $I_d$ < $I_l$.   For a given value of the thresholds and of the $n^{\gamma}_{\alpha l}$  we can compute the Nash equilibrium as  in our original approach, and we can then perform a gradient descent on these parameters to reach their optimal value, i.e. the optimal lockdown policy. 

We show on figure ~\ref{fig.SO_constraints} the numerical simulation for the societal optimum and for the optimal lockdown policy, with the same parameters as Fig.~\ref{fig.MFG-free} and $r_I = 1$, giving for the optimal lockdown policy  $I_l = 0.1 = I_{sat}$, $I_d = 4.10^{-4}$, and $\sigma =0.39$. 
For the societal optimum the cooperation appears clearly : at the epidemic peak there is a mutual action of all individuals to limit simultaneously their contacts, especially in community and households where adults and young people make efforts in order to limit the number of infected retired people, even if the efforts in households are costly. On the other hand, less efforts are made in schools or in workplaces because this affect less the retired people. These combination of efforts leads to a very low cost for the entire society. 

For the Nash equilibrium under optimal constraints we see on figure ~\ref{fig.SO_constraints} that since the lockdown is imposed in all settings, we obtain a situation where adults and young people make almost the same efforts everywhere while retired people go beyond  the lockdown level to achieve further protection. This lockdown has a strong effect on the epidemic but lack the coordination of the optimum societal case. This leads to a number of infected adults and young which is lower than the {\em ``societal optimum''} while it is higher for retired people.

To conclude it might be useful to introduce a ``figure of merite'' of a given policy $\mathcal{P}$
\begin{equation}
  M(\mathcal{P}) = \frac{C_{\rm glob}(\mathcal{P})  - C_{\rm glob}(\mbox{societal optima}) }{C_{\rm glob}(\mbox{business as usual})  - C_{\rm glob}(\mbox{societal optima})} \; , 
\end{equation}
which is thus such that $M(\mathcal{P})$ is 0 if $\mathcal{P}$ is the societal optima and 1 if $\mathcal{P}$ is the ``business as usual'' strategy for which no adjustment is made to the contacts between agents.  On that scale, we see that $M(\mathcal{P}) = 0.12$ for the unconstrained Nash equilibrium and 0.06 for the optimal lockdown policy, but non-optimal lockdown policies are most of the time less effective than the unconstrained Nash, and can have typically a $M$ of order $0.5$.

Although these numbers apply obviously only to the specific model and to the specific set of parameters we have used as an illustration here, there is no doubt that the qualitative features observed are very general in nature.  Namely, the Nash equilibrium is already a very significant improvement with respect to the ``business as usual approach'', and if on the one hand an optimized lockdown strategy can close further the gap toward the societal, sub-optimal lockdown strategies can actually ``degrade'' the situation with respect to the basic Nash equilibrium.

As a final remark, we stress that it should not be assumed, and we certainly do not imply here, that the Nash equilibrium is the ``natural outcome'' of the epidemic process that would be reached  in the absence of any public policy.  Indeed, our model assumes that the agents possess both perfect information and the technical resources to compute the Nash equilibrium, which we cannot expect them to have in practice. On the other hand gathering this information and developing the technical tools to compute the Nash equilibrium appears like a reachable goal for a centralized public agency.  If enough trust is build between the government and the individual agents, making public that information can be enough to coordinate the ensemble of agents around the Nash equilibrium.  This, as well as the optimization of lockdown or similar policies improving on the basic Nash equilibrium requires to develop the necessary conceptual tools.  We hope this work provides a useful step in that direction.

\bibliography{biblio}
\bibliographystyle{plain}

\end{document}